\documentclass[11pt]{article}

\usepackage[margin=2cm]{geometry}
\usepackage{amsmath, amssymb, amsthm}
\usepackage{graphicx}
\usepackage{subcaption}
\usepackage{hyperref}
\usepackage{booktabs}
\usepackage{xcolor}
\usepackage{nicefrac}
\usepackage{bm,cite}
\usepackage{microtype}
\usepackage{float}

\usepackage{doi}
\hypersetup{
  colorlinks   = true,
  linkcolor    = blue!60!black,
  citecolor    = blue!60!black,
  urlcolor     = blue!60!black,
  breaklinks   = true,
  pdfborder    = {0 0 0}
}

\newcommand{\avg}[1]{\langle #1 \rangle}
\newcommand{\fgdist}{p_{\text{FG}}}

\newcommand{\ee}{\mathrm{e}}
\newcommand{\dd}{\mathrm{d}}
\newcommand{\erf}{\mathrm{erf}}

\title{Statistical Mechanics of a Quantum Harmonic Oscillator\\
       with Folded Gaussian Frequency}

\author{Liu Zhao\\[2pt]
School of Physics, Nankai University, Tianjin 300071, China\\
email: \texttt{[lzhao@nankai.edu.cn]}}

\begin{document}
\maketitle

% ============================================================
\begin{abstract}
A self-contained statistical-mechanics treatment of a single 
quantum harmonic oscillator is presented, whose frequency $\omega$ is drawn from a 
folded Gaussian distribution: $\omega=|\xi|$ with $\xi\sim\mathcal{N}(\mu,\sigma^2)$.
The exact integral representations for the partition function, internal energy, 
free energy, heat capacity, and entropy are derived, and analytic 
approximations are given in two complementary limits---small variance ($\sigma\ll\mu$) 
via a cumulant expansion, and the zero-center case ($\mu=0$) via low-frequency 
asymptotic analysis. The model is extended to $N$ independent oscillators, 
where the heat capacity is shown to be extensive with self-averaging fluctuations 
$\propto N^{-1/2}$, and finally to a disordered oscillator lattice, where 
the folded-Gaussian kink at $\omega=0$ produces a soft-mode infrared tail.
For a single isolated oscillator with $\mu=0$, both $C$ and $S$ 
vanish linearly at low $T$. In the lattice case, the van Hove factor converts 
this to a $T^d$ power law. The oft-quoted ``third-law violation'' for disordered 
phonons is here shown to be a spectral property---the absence of an energy gap and 
a power-law freeze-out---driven by the single-site distribution kink rather than 
by a genuine Lifshitz tail (which requires rare large-scale spatial fluctuations). 
The folded Gaussian thus serves as a minimal benchmark for soft-mode disorder 
thermodynamics.
\end{abstract}

% ============================================================
\section{Introduction}
\label{sec:intro}

The harmonic oscillator is the paradigmatic building block of quantum 
statistical mechanics: its spectrum is exactly solvable, its canonical 
partition function is the starting point for phonon theory, and its 
low-temperature behavior encodes the third law of thermodynamics.
Introducing randomness into the oscillator frequency transforms this 
elementary model into a window onto disordered condensed matter, 
connecting to disordered oscillator lattices ~\cite{matsuda1970} 
and, more distantly, the Lifshitz-tail phenomenology of random 
Schrödinger / phonon systems ~\cite{lifshitz1964,gredeskul1988}. 
The model to be presented in this work, however, produces a 
\emph{soft-mode kink tail} rather than a genuine Lifshitz tail: the infrared 
weight arises from the non-analyticity of $\omega=|\xi|$ at $\omega=0$, 
not from rare large-scale spatial fluctuations.

Consider the simplest such disorder: a single quantum harmonic oscillator
\begin{equation}
\hat H(\omega) = \frac{\hat p^2}{2m} + \frac12 m\omega^2\hat x^2,
\end{equation}
with frequency $\omega=|\xi|$ drawn from a folded Gaussian distribution,
\begin{equation}
\xi \sim \mathcal{N}(\mu,\sigma^2), \qquad \omega = |\xi| \ge 0.
\end{equation}
The absolute value ensures $\omega>0$ more naturally than a truncated Gaussian and, 
crucially, creates a non-analytic kink at the origin that generates a soft-mode 
tail with distinctive thermodynamic consequences.

\paragraph{What is already known.}
Static disorder in oscillator systems \cite{dyson1953,matsuda1970} 
has been studied in several contexts: Anderson localization~\cite{anderson1958}, 
disordered oscillator chains (entanglement scaling and area 
laws)~\cite{audenaert2002,eisert2010,plenio2005}, and the Lifshitz-tail phenomenology 
of random potentials~\cite{lifshitz1964}. The folded Gaussian is deliberately 
minimal: it guarantees positivity, preserves a Gaussian parent,
and produces a \emph{soft-mode kink tail} that is different from a genuine Lifshitz 
tail. The folded Gaussian distribution itself is a well-studied object 
in statistics~\cite{tsagris2014}.
However, the specific combination of a folded Gaussian frequency with full 
canonical averaging---and the resulting soft-mode phenomenology---has not 
been isolated in the literature.

\paragraph{This paper.}
I provide exact integral representations, analytic approximations, 
finite-$N$ scaling laws, and a lattice extension.
A key result is the low-temperature behavior for $\mu=0$: the heat capacity 
of a single oscillator is \emph{linear} in $T$, $C\propto T$, and the third law 
is satisfied; in a lattice the van Hove factor $\omega^{d-1}$ converts this to 
$C \propto N T^d$. The structure is as follows.
Section~\ref{sec:fg} defines the folded Gaussian distribution and its moments.
Section~\ref{sec:single} gives the exact thermodynamics of a single oscillator.
Section~\ref{sec:ensemble} introduces the ensemble average, involving 
the small-$\sigma$ cumulant expansion and the analysis of the $\mu=0$ limit. 
Section~\ref{sec:N} extends to $N$ independent oscillators.
Section~\ref{sec:lattice} embeds the model in a disordered lattice.
Section~\ref{sec:third-law} clarifies the third law: satisfied for a single 
oscillator, with the lattice difference being spectral rather than entropic. 
Section~\ref{sec:conclusions} summarizes.

% ============================================================
\section{The Folded Gaussian Distribution}
\label{sec:fg}

The folded Gaussian probability density for $\omega\ge0$ is
\begin{equation}\label{eq:fg-pdf}
\fgdist(\omega;\mu,\sigma) = \frac{1}{\sqrt{2\pi}\,\sigma}
\left[
\exp\!\left(-\frac{(\omega-\mu)^2}{2\sigma^2}\right)
+ \exp\!\left(-\frac{(\omega+\mu)^2}{2\sigma^2}\right)
\right],
\end{equation}
with $\displaystyle\int_0^\infty \fgdist(\omega)\,\dd\omega = 1$.
Here $\mu$ has units of $\mathrm{s^{-1}}$ and $\sigma$ the same, 
so that $\fgdist$ has units of $\mathrm{s}$.

The first two moments are~\cite{tsagris2014}
\begin{align}
\avg{\omega} &= \mu\left[1-2\Phi\!\left(-\frac{\mu}{\sigma}\right)\right]
+ \sigma\sqrt{\frac{2}{\pi}}\exp\!\left(-\frac{\mu^2}{2\sigma^2}\right), \label{eq:moment1}\\[4pt]
\avg{\omega^2} &= \mu^2+\sigma^2, \label{eq:moment2}
\end{align}
where $\Phi(z)=\frac12[1+\erf(z/\sqrt2)]$ is the standard normal cumulative 
distribution function.

Two limits are especially relevant for the physics that follows.
\begin{itemize}
\item \textbf{Sharp frequency} ($\mu\gg\sigma$): $\avg{\omega}\simeq\mu$, 
$\mathrm{Var}(\omega)\simeq\sigma^2$, and the distribution approaches a narrow 
Gaussian centered at $\mu$.
This is the regime where the cumulant expansion of Section~\ref{sec:cumulant} applies.
\item \textbf{Zero center} ($\mu=0$): the distribution becomes half-normal,
$\fgdist(\omega;0,\sigma)=\sqrt{2/\pi}\,\sigma^{-1}\ee^{-\omega^2/(2\sigma^2)}$,
with $\avg{\omega}=\sigma\sqrt{2/\pi}$.
The density at $\omega=0$ is finite and nonzero---this is the soft-mode tail.
For a single oscillator this tail produces linear low-$T$ 
thermodynamics (Section~\ref{sec:mu0}); in a lattice it produces power-law 
behavior (Section~\ref{sec:lattice}).
\end{itemize}

\begin{figure}[H]
\centering
\includegraphics[width=0.6\textwidth]{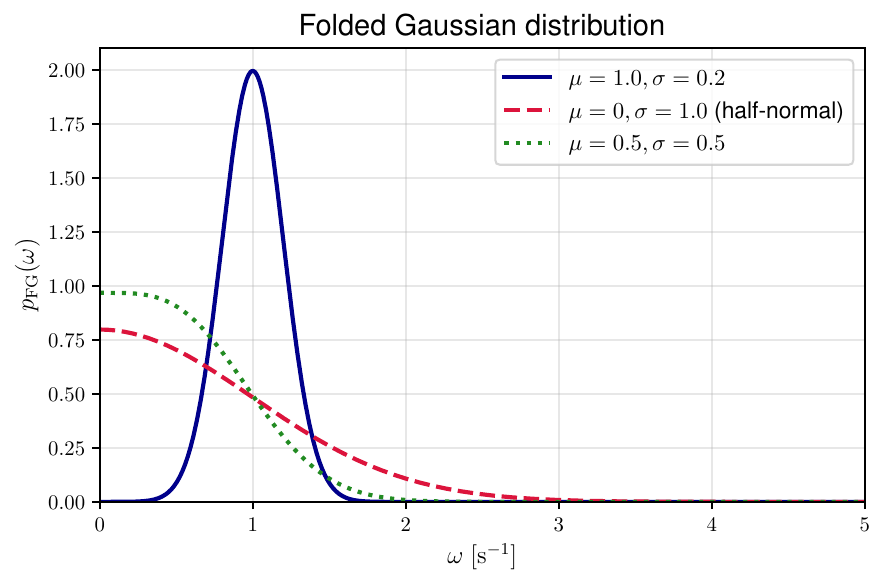}
\caption{Folded Gaussian probability density $\fgdist(\omega;\mu,\sigma)$ for 
three parameter sets.
The solid blue curve ($\mu=1.0\;\mathrm{s^{-1}}$, $\sigma=0.2\;\mathrm{s^{-1}}$) 
is sharply peaked near $\omega=\mu$, justifying the cumulant expansion of 
Sec.~\ref{sec:cumulant}.
The dashed red curve ($\mu=0$, $\sigma=1.0\;\mathrm{s^{-1}}$) is the 
half-normal distribution; its finite, nonzero value at $\omega=0$ is the soft-mode tail.
The dotted green curve ($\mu=0.5\;\mathrm{s^{-1}}$, $\sigma=0.5\;\mathrm{s^{-1}}$) 
interpolates between the two limits.}
\label{fig:fg-dist}
\end{figure}

% ============================================================
\section{Single-Mode Exact Thermodynamics}
\label{sec:single}

For a fixed frequency $\omega$, the canonical partition function is
\begin{equation}\label{eq:Z}
Z(\beta,\omega) = \frac{\ee^{-\beta\hbar\omega/2}}{1-\ee^{-\beta\hbar\omega}}
= \frac{1}{2\sinh(\beta\hbar\omega/2)},
\end{equation}
with $\beta=1/(k_B T)$.
The key thermodynamic quantities follow:
\begin{align}
F(\beta,\omega) &= \frac{\hbar\omega}{2} 
+ \beta^{-1}\ln\!\left(1-\ee^{-\beta\hbar\omega}\right), \label{eq:F-exact}\\[4pt]
U(\beta,\omega) &= \frac{\hbar\omega}{2} + \frac{\hbar\omega}{\ee^{\beta\hbar\omega}-1}, 
\label{eq:U-exact}\\[4pt]
C(\beta,\omega) &= k_B\,(\beta\hbar\omega)^2
\frac{\ee^{\beta\hbar\omega}}{(\ee^{\beta\hbar\omega}-1)^2}
=k_B\,\left[\frac{\beta\hbar\omega}{2\sinh(\beta\hbar\omega/2)}\right]^2, 
\label{eq:C-exact}\\[4pt]
S(\beta,\omega) &= k_B\beta \,[U(\beta,\omega) - F(\beta,\omega)].
\end{align}

\paragraph{Limits.} At high temperature ($k_B T\gg \hbar\omega$) equipartition 
is recovered: $U\to k_B T$, $C\to k_B$, and the entropy reduces to the 
classical harmonic-oscillator form
\begin{equation}
S(\beta,\omega)\simeq k_B\left[1+\ln\left(\frac{k_B T}{\hbar\omega}\right)
+\frac{1}{24}\left(\frac{\hbar\omega}{k_B T}\right)^2\right],
\end{equation}
reflecting the logarithmic growth of phase space with temperature. 
At low temperature ($k_B T\ll \hbar\omega$) the system freezes exponentially: 
$U\simeq\hbar\omega/2+\hbar\omega \ee^{-\beta\hbar\omega}$, 
$C\propto(\beta\hbar\omega)^2\ee^{-\beta\hbar\omega}$, and the entropy is 
exponentially suppressed,
\begin{equation}
S(\beta,\omega)\simeq k_B\left(\frac{\beta\hbar\omega}{2}\right)^2
\ee^{-\beta\hbar\omega}\to0,
\end{equation}
verifying the third law for a fixed-frequency oscillator.

% ============================================================
\section{Ensemble-Averaged Quantities}
\label{sec:ensemble}

Averaging over the folded Gaussian ensemble gives
\begin{equation}\label{eq:avg-def}
\avg{\mathcal{O}(\beta)} = \int_0^\infty \dd\omega\,
\fgdist(\omega;\mu,\sigma)\,\mathcal{O}(\beta,\omega).
\end{equation}
Concretely,
\begin{align}
\avg{F(\beta)} &= \int_0^\infty \dd\omega\,\fgdist(\omega)\left[\frac{\hbar\omega}{2} 
+ \beta^{-1}\ln\!\left(1-\ee^{-\beta\hbar\omega}\right)\right], \label{eq:avg-F}\\[4pt]
\avg{U(\beta)} &= \int_0^\infty \dd\omega\,\fgdist(\omega)\left[\frac{\hbar\omega}{2} 
+ \frac{\hbar\omega}{\ee^{\beta\hbar\omega}-1}\right], \label{eq:avg-U}\\[4pt]
\avg{C(\beta)} &= k_B\int_0^\infty \dd\omega\,\fgdist(\omega)\,
\left[\frac{\beta\hbar\omega}{2\sinh(\beta\hbar\omega/2)}\right]^2. 
\label{eq:avg-C}
\end{align}
These integrals are smooth and numerically stable but have no elementary 
closed form. Below I will consider two special limits $\mu\gg\sigma$ and $\mu=0$ 
and employ different approximations.

% ============================================================

\subsection{Small-$\sigma$ Cumulant Expansion ($\mu\gg\sigma$)}
\label{sec:cumulant}
When the center $\mu$ is much larger than the width $\sigma$, the folded Gaussian 
distribution is sharply peaked near $\omega=\mu$, and the probability weight 
for $|\omega-\mu|\gtrsim\sigma$ is exponentially suppressed. Any thermodynamic 
observable $g(\beta,\omega)$ depending parametrically on the frequency can be
expanded around $\omega=\mu$ to the second order, and, by taking the average over the 
folded Gaussian ensemble, a controlled cumulant approximation arises,
\begin{align}
\langle g(\beta,\omega)\rangle 
&=\int_0^\infty \dd\omega\,\fgdist(\omega) g(\beta,\omega)
\simeq g(\beta,\mu) +\frac{1}{2} 
\frac{\partial^2 g(\beta,\omega)}{\partial\omega^2}\bigg|_{\omega=\mu} \,
\langle (\omega-\mu)^2 \rangle\nonumber\\
&\simeq g(\beta,\mu) +\frac{\sigma^2}{2} 
\frac{\partial^2 g(\beta,\omega)}{\partial\omega^2}\bigg|_{\omega=\mu},
\label{eq:cumulant}
\end{align}
where, for $\mu\gg\sigma$, the mean shift vanishes to leading order, 
$\langle\omega\rangle=\mu+\mathcal{O}(\sigma^2/\mu)$, and the variance saturates 
to $\langle(\omega-\mu)^2\rangle\simeq\sigma^2$ (see Eq.~\eqref{eq:moment2}). 

\subsubsection{Free Energy}
From Eq.~\eqref{eq:F-exact},
\begin{align}
\frac{\partial^2 F(\beta,\omega)}{\partial\omega^2}\bigg|_{\omega=\mu} 
&= -\frac{\beta\hbar^2 \ee^{x_0}}{(\ee^{x_0}-1)^2},\qquad x_0\equiv\beta\hbar\mu.
\end{align}
Substituting into Eq.~\eqref{eq:cumulant} gives the ensemble-averaged free energy
\begin{equation}
\langle F(\beta)\rangle \simeq \frac{\hbar\mu}{2}+\beta^{-1} \ln\!\left(1-\ee^{-x_0}\right)
-\frac{\sigma^2}{2}\,\frac{\beta\hbar^2 \,\ee^{x_0}}{(\ee^{x_0}-1)^2},
\label{eq:avg-F-cumulant}
\end{equation}
The negative correction confirms that frequency fluctuations lower the free energy: 
a pure entropy-of-disorder effect.

\subsubsection{Internal Energy}
Differentiating Eq.~\eqref{eq:U-exact} twice with respect to $\omega$ yields
\begin{align}
\frac{\partial^2 U(\beta,\omega)}{\partial\omega^2}\bigg|_{\omega=\mu} 
&= \beta\hbar^2\,\frac{\ee^{x_0}\big[x_0(\ee^{x_0}+1)-2(\ee^{x_0}-1)\big]}
{(\ee^{x_0}-1)^3}.
\end{align}
The cumulant approximation is therefore
\begin{equation}
\langle U(\beta)\rangle \simeq \frac{\hbar\mu}{2}+\frac{\hbar\mu}{\ee^{x_0}-1}
+\frac{\sigma^2}{2}\,\beta\hbar^2\,\frac{\ee^{x_0}\big[x_0(\ee^{x_0}+1)
-2(\ee^{x_0}-1)\big]}{(\ee^{x_0}-1)^3}.
\label{eq:avg-U-cumulant}
\end{equation}

\subsubsection{Heat Capacity}
\begin{figure}[htbp]
\begin{center}
\includegraphics[width=\textwidth]{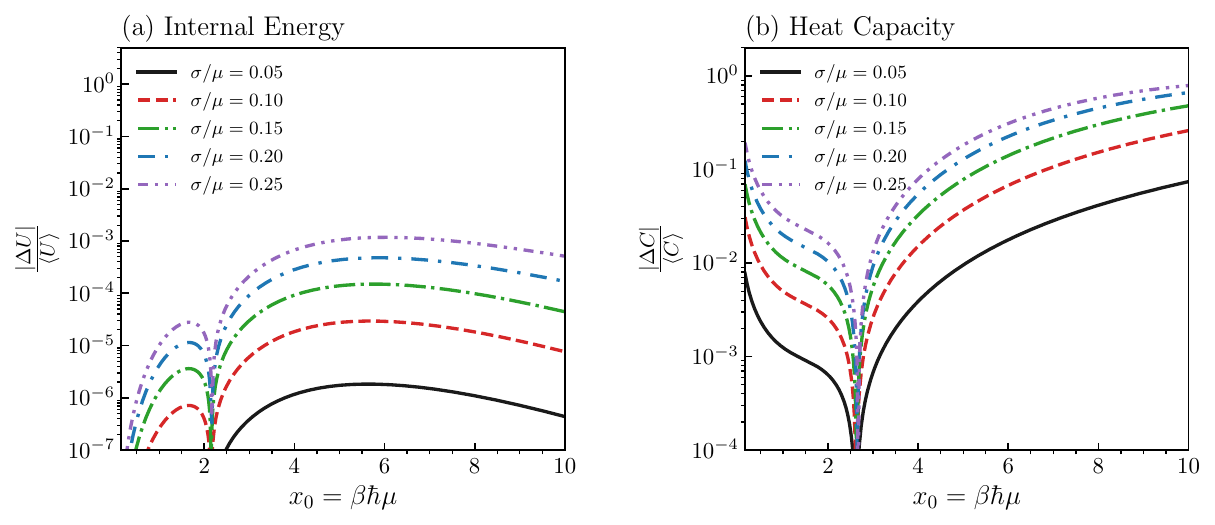}
\end{center}
\caption{Relative errors of the cumulant expansion: (a) internal energy; 
(b) heat capacity. The expansion retains only the variance term. For $U$, 
errors remain below $10^{-3}$ for $\sigma/\mu\leq 0.20$; 
for $C$, deviation is stronger and errors can reach $\sim 80\%$ at large $\sigma/\mu$, 
confirming cumulant truncation is insufficient for heat capacity in the 
high-$T$ regime.}
\label{fig:rel-error}
\end{figure}

A similar calculation gives
\begin{equation}
\frac{\partial^2 C(\beta,\omega)}{\partial\omega^2}\bigg|_{\omega=\mu}
= k_B(\beta\hbar)^2 \,\frac{\ee^{x_0}\big[x_0^2\ee^{x_0}+4x_0\ee^{x_0}
+4\ee^{x_0}-4\big]}{8(\ee^{x_0}-1)^4}.
\label{eq:d2C}
\end{equation}
The ensemble-averaged heat capacity reads
\begin{equation}
\langle C(\beta)\rangle \simeq k_B \left[\frac{x_0^2 \ee^{x_0}}{(\ee^{x_0}-1)^2}
+\frac{1}{16}(\beta\hbar\sigma)^2\,\frac{\ee^{x_0}\big[x_0^2 \ee^{x_0}
+4x_0 \ee^{x_0}+4\ee^{x_0}-4\big]}{(\ee^{x_0}-1)^4}\right].
\label{eq:avg-C-cumulant}
\end{equation}
Figure~\ref{fig:rel-error}(a)(b) shows the relative error of the internal energy
and heat capacity, 
validating the cumulant expansion for $U$ but not for $C$. 
The reason that the relative error of $C$ is more difficult to control lies in that 
$C$ has stronger nonlinearity in $\omega$, consequently the cumulant expansion 
for $C$ is uncontrollable, even with the inclusion of higher order corrections.

\subsection{The $\mu=0$ Limit: Half-Normal Frequency Distribution}
\label{sec:mu0}
Setting $\mu=0$ reduces the folded Gaussian to the half-normal distribution,
\begin{equation}
p_{\mathrm{FG}}(\omega;0,\sigma)=\sqrt{\frac{2}{\pi}}\frac{1}{\sigma}
\exp\left(-\frac{\omega^2}{2\sigma^2}\right),\qquad\omega\ge0,
\label{eq:half-normal}
\end{equation}
with mean $\langle\omega\rangle=\sigma\sqrt{2/\pi}$ and finite probability density at 
the origin $p_{\mathrm{FG}}(0;0,\sigma)=\sqrt{2/\pi}/\sigma>0$. 
The zero-temperature limit retains only the zero-point energy:
\begin{equation}
\langle U\rangle_{\mu=0}(T\to0)=\frac{\hbar}{2}\int_0^\infty 
p_{FG}(\omega;0,\sigma)\,\omega\,\dd\omega=\frac{\hbar\sigma}{\sqrt{2\pi}},
\end{equation}
confirming a unique vacuum ground state for the single oscillator.

\subsubsection{Low-Temperature Heat Capacity}

Substituting Eq.~\eqref{eq:half-normal} into Eq.~\eqref{eq:avg-C} gives
\begin{equation}
\langle C(\beta)\rangle_{\mu=0}
=k_B\sqrt{\frac{2}{\pi}}\frac{1}{\sigma}\int_0^\infty \dd\omega\,
\exp\left(-\frac{\omega^2}{2\sigma^2}\right)\left[\frac{\beta\hbar\omega}
{2\sinh(\beta\hbar\omega/2)}\right]^2.
\label{eq:C-mu0}
\end{equation}
At $k_BT\ll\hbar\sigma$ the integral is dominated by $x=\beta\hbar\omega\lesssim1$. 
For these modes $\beta\hbar\omega\le\mathcal{O}(1)$, 
the Bose factor reduces to $[x/(2\sinh(x/2))]^2\simeq 1+\mathcal{O}(x^2)$, 
while the Gaussian density varies negligibly from its origin value $p_{\mathrm{FG}}(0)$. 
Splitting the integral at $\omega_*=k_BT/\hbar$, the tail $\omega>\omega_*$ is 
exponentially suppressed by both the gapped Bose factor and the Gaussian weight, 
and contributes only $\mathcal{O}(\ee^{-1})$ corrections to the leading order. 
Keeping the soft-mode window and expanding to leading order,
a controlled estimate gives
\begin{align}
\langle C(T)\rangle_{\mu=0}&\simeq 
k_B\sqrt{\frac{2}{\pi}}\frac{k_B T}{\hbar\sigma}
\int_0^\infty \dd u\,\frac{u^2}{4\sinh^2(u/2)}\nonumber\\
&= \frac{\sqrt{2}\pi^{3/2}}{3}\,k_B\,\left(\frac{k_B T}{\hbar\sigma}\right)
\approx 2.63\,k_B\,\frac{k_B T}{\hbar\sigma},
\qquad k_BT\ll\hbar\sigma.
\label{eq:C-mu0-final}
\end{align}

\subsubsection{Low-Temperature Entropy: Rigorous Derivation}
The entropy is $\langle S\rangle=T^{-1}(\langle U\rangle-\langle F\rangle)$. 
Using the half-normal density and exact single-oscillator results, one gets 
\begin{equation}
\langle S(T)\rangle =k_B\sqrt{\frac{2}{\pi}}\frac{k_B T}{\hbar\sigma}
\int_0^\infty \dd u\,\ee^{-u^2/(2y^2)}\left[\frac{u}{\ee^u-1}-\ln(1-\ee^{-u})\right],
\quad y=\frac{\hbar\sigma}{k_BT}\gg1,
\label{eq:S-mu0-integral}
\end{equation}
where the dimensionless frequency $u$ is defined via $u=\beta\hbar\omega$.

The integrand $g(u)=u/(\ee^u-1)-\ln(1-\ee^{-u})$ has a logarithmic singularity 
at $u\to0$, but $\displaystyle\int_0^\epsilon\ln u\,\dd u=\epsilon\ln\epsilon-\epsilon$ 
is finite as $\epsilon\to0^+$, so $g(u)$ is Lebesgue-integrable at the origin. 
For $u\gg1$, $g(u)\sim u \ee^{-u}$ ensures UV convergence. Thus the dimensionless 
integral is absolutely convergent for any $y>0$.

In the low-temperature limit $y\gg1$, the prefactor $\ee^{-u^2/(2y^2)}\to1$ 
uniformly for $u=\mathcal{O}(1)$, and the integral reduces to
$$
\int_0^\infty\left[\frac{u}{\ee^u-1}-\ln(1-\ee^{-u})\right]\dd u
=\sum_{n=1}^\infty\int_0^\infty\left(\frac{u}{\ee^u-1}+\frac{\ee^{-nu}}{n}\right)\dd u
=\frac{\pi^2}{3}.
$$
Hence
\begin{equation}
\langle S(T)\rangle_{\mu=0}
\simeq \frac{\sqrt{2}\pi^{3/2}}{3}\,k_B\,\left(\frac{k_B T}{\hbar\sigma}\right)
\approx 2.63\,k_B\,\frac{k_B T}{\hbar\sigma},\qquad T\to0.
\label{eq:S-mu0-final}
\end{equation}
The entropy vanishes linearly with $T$, thus the third law still holds for a single 
oscillator with frequency disorder.

\section{$N$ Independent Oscillators}
\label{sec:N}
Consider $N$ uncoupled oscillators with iid folded-Gaussian frequencies 
$\omega_i=|\xi_i|$, $\xi_i\sim\mathcal N(\mu,\sigma^2)$. For a fixed frequency 
configuration $\{\omega_i\}$, the total partition function factorizes:
\begin{equation}
\mathcal Z_N(\beta;\{\omega_i\})=\prod_{i=1}^N Z_1(\beta,\omega_i),\qquad 
Z_1(\beta,\omega)=\frac{1}{2\sinh(\beta\hbar\omega/2)}.
\end{equation}
Averaging over the disorder ensemble and using independence,
\begin{equation}
\langle\mathcal Z_N\rangle=\Big(\int_0^\infty p_{\mathrm{FG}}(\omega;\mu,\sigma)
Z_1(\beta,\omega)\,\dd\omega\Big)^N\equiv\langle Z_1\rangle^N,
\end{equation}
and the total internal energy and heat capacity are sums of single-oscillator contributions,
\begin{equation}
\langle U_N\rangle =\sum_{i=1}^N \langle u(\beta,\omega_i)\rangle,\qquad 
\langle C_N\rangle =\sum_{i=1}^N \langle c(\beta,\omega_i)\rangle,
\end{equation}
with $u(\beta,\omega)=\hbar\omega/2+\hbar\omega/(\ee^{\beta\hbar\omega}-1)$ and 
$c(\beta,\omega)=k_B[\beta\hbar\omega/(2\sinh(\beta\hbar\omega/2))]^2$. 
This gives
\begin{equation}
\frac{\langle U_N\rangle}{N}=\langle u\rangle,\qquad \frac{\langle C_N\rangle}{N}
=\langle c\rangle,
\end{equation}
indicating extensivity in the $\omega$-ensemble avarage.

To characterize sample-to-sample fluctuations, treat $X_i=c(\beta,\omega_i)$ as 
iid copies with mean $m_c=\langle c\rangle$ and variance 
$s_c^2=\langle c^2\rangle-\langle c\rangle^2$ (both finite for the folded Gaussian). 
Then
\begin{equation}
\mathrm{Var}(C_N)=N s_c^2,\qquad
\frac{\sqrt{\mathrm{Var}(C_N)}}{\langle C_N\rangle}=\frac{s_c}{m_c}\frac{1}{\sqrt{N}}.
\end{equation}
Thus the relative fluctuation of $C_N/N$ decays as $N^{-1/2}$: the system is 
self-averaging. By the central limit theorem,
\begin{equation}
\frac{C_N-Nm_c}{\sqrt{N}\,s_c}\xrightarrow{d}\mathcal N(0,1),
\end{equation}
i.e.\ the sample-to-sample heat capacity per oscillator approaches a Gaussian 
distribution with width $\propto N^{-1/2}$. The same reasoning applies to $U_N/N$ and to 
the total free energy $F_N=k_BT\sum_i\ln Z_1(\beta,\omega_i)$. The $N^{-1/2}$ 
scaling is the standard signature of self-averaging in extensive quantities 
built from iid disorder.

\section{Disordered Lattice}
\label{sec:lattice}

\paragraph{Model Hamiltonian.}
Consider a $d$-dimensional hypercubic lattice with $N$ sites, with
scalar displacements $u_i$, mass $m$, nearest-neighbour spring constant $\kappa$, 
and $N$ independent on-site frequency disorder $\omega_i$. The Hamiltonian reads
\begin{equation}
\hat H=\sum_{i=1}^N\left[\frac{\hat p_i^2}{2m}+\frac12 m\omega_i^2 u_i^2\right]
+\frac{\kappa}{2}\sum_{\langle ij\rangle}(u_i-u_j)^2,\qquad\omega_i=|\xi_i|,\;
\xi_i\sim\mathcal N(\mu,\sigma^2).
\label{eq:lattice-H}
\end{equation}
The last term provides acoustic propagation; the term $\propto\omega_i^2$ 
encodes on-site frequency disorder.

\paragraph{Disorder-averaged density of states (DOS).}
In the clean limit ($\omega_i\equiv0$) the model reduces to the standard scalar 
phonon chain/lattice. Fourier transforming $\tilde u_{\mathbf k}
=N^{-1/2}\sum_i u_i \ee^{-i\mathbf k\cdot\mathbf r_i}$ gives the dispersion
$$
\Omega_{\mathbf k}=2\sqrt{\frac{\kappa}{m}}\left|\sin\left(\frac{ka}{2}\right)\right|,
\quad k_\alpha=\frac{2\pi n_\alpha}{L_\alpha},
$$
with low-energy linear form $\Omega_{\mathbf k}\simeq v_s|\mathbf k|$, 
$v_s=a\sqrt{\kappa/m}$. The clean-lattice DOS per site is
$$
\rho_0(\omega)=\frac{V_d}{(2\pi)^d}\int \dd^dk\,\delta(\omega-\Omega_{\mathbf k})
\simeq \frac{d\,\omega^{d-1}}{v_s^d\,\Gamma(d/2+1)2^{d-1}\pi^{d/2}}
\equiv A_d\,\omega^{d-1},\quad\omega\to0,
$$
exhibiting the usual van Hove power-law singularity $\rho_0(\omega)\sim\omega^{d-1}$ 
at the acoustic band edge. 

When on-site frequencies are drawn independently 
from a folded Gaussian, the eigenmodes hybridize weakly, and to leading order 
the average DOS factorizes:
\begin{equation}
\bar\rho(\omega)=A_d\,\omega^{d-1}\,p_{\mathrm{FG}}(\omega;\mu,\sigma),
\label{eq:lattice-DOS}
\end{equation}
i.e.\ each van Hove shell is weighted by the single-site frequency density.
For $\mu=0$, $p_{\mathrm{FG}}(0)=\sqrt{2/\pi}/\sigma>0$, so 
$\bar\rho(\omega)\sim\omega^{d-1}$ at low $\omega$.
Eq.~\eqref{eq:lattice-DOS} is exact 
for decoupled oscillators and captures the infrared soft-mode tail of 
the disordered lattice. For \emph{mass} or \emph{spring-constant} disorder, 
the same shell picture leads to genuine Lifshitz tails with log-periodic 
prefactors~\cite{nieuwenhuizen1987}, as reviewed for chains in \cite{matsuda1970}; 
the folded Gaussian replaces those 
large-deviation singularities by a plain $\omega^{d-1}$ kink weight.

\paragraph{Ensemble-averaged heat capacity.}
The total heat capacity is the disorder-averaged sum over normal modes:
\begin{align}
\langle C(T)\rangle &=k_B\int_0^{\omega_D} \dd\omega\,\bar\rho(\omega)
\left[\frac{\beta\hbar\omega}{2\sinh(\beta\hbar\omega/2)}\right]^2\nonumber\\
&\simeq k_B A_d\int_0^\infty \dd\omega\,\omega^{d-1}p_{\mathrm{FG}}(\omega;\mu,\sigma)
\,\left[\frac{\beta\hbar\omega}{2\sinh(\beta\hbar\omega/2)}\right]^2,
\label{eq:CV-lattice}
\end{align}
where $\omega_D\sim v_s\pi/a$ is the Debye cutoff (irrelevant for low $T$).

\paragraph{Case I: $\mu>0$ (gapped tail).}
For $k_BT\ll\hbar\mu$, the folded Gaussian is approximately constant over 
$[0,k_BT/\hbar]$ and exponentially suppressed for $\omega\ll\mu$: 
$\displaystyle p_{\mathrm{FG}}(\omega;\mu,\sigma)\simeq\frac{2}{\sqrt{2\pi}\sigma}
\ee^{-\mu^2/(2\sigma^2)}\ee^{\mu\omega/\sigma^2}$. The Bose kernel is gapped, 
so the low-$T$ contribution comes from the lowest van Hove shell:
\begin{equation}
\langle C(T)\rangle \simeq k_B A_d\frac{2\ee^{-\mu^2/(2\sigma^2)}}{\sqrt{2\pi}\sigma}
\int_0^\infty \dd\omega\,\omega^{d+1}\ee^{-\hbar\omega/k_BT}\propto N\,T^d 
\ee^{-\hbar\mu/k_BT},
\end{equation}
the standard Debye exponential freeze-out multiplied by a disorder prefactor.

\paragraph{Case II: $\mu=0$ (soft-mode lattice).}
With $\mu=0$, $\displaystyle p_{\mathrm{FG}}(\omega;0,\sigma)
=\sqrt{2/\pi}\sigma^{-1}\ee^{-\omega^2/(2\sigma^2)}\to p_0\equiv\sqrt{2/\pi}/\sigma$ 
as $\omega\to0$. At $k_BT\ll\hbar\sigma$, the relevant frequencies satisfy 
$\beta\hbar\omega\ll1$, so the Gaussian cutoff is irrelevant over the soft window. 
The integral reduces to a pure phase-space power law:
\begin{align}
\langle C(T)\rangle &\simeq k_B A_d\,p_0\int_0^{\omega_c}\omega^{d-1}\dd\omega
\left[1+\mathcal{O}\!\left(\frac{k_BT}{\hbar\sigma}\right)^2\right] \nonumber\\
&=k_B A_d\,p_0\frac{(k_BT/\hbar)^d}{d}\propto N\left(\frac{k_BT}{\hbar\sigma}\right)^d,
\qquad T\to0.
\label{eq:CV-mu0-lattice}
\end{align}
The heat capacity never freezes exponentially; instead it follows a $T^d$ power law with an extensive prefactor.

\paragraph{Third-law interpretation.}
For a single oscillator (Section~\ref{sec:mu0}) the phase-space volume is 
$\sim\omega_c\propto T$, 
giving $\langle C\rangle \propto T$ and $\langle S\rangle\to0$ smoothly. 
In the lattice, the extensive van Hove factor 
$\omega^{d-1}$ converts the single-particle $T$ law into $T^d$ with 
total $\langle C(T) \rangle \propto N T^d$. 
Strictly speaking $\langle S(T) \rangle/N\propto T^d\to0$ (Nernst form holds), 
but the absence of an exponential gap and the macroscopic weight of soft modes 
is sometimes called a ``Lifshitz-tail signature'' in the disordered-phonon literature
\cite{lifshitz1964,matsuda1970,nieuwenhuizen1987}, though strictly it is the 
folded-Gaussian kink tail discussed above. The lattice \emph{thermodynamically 
softens} for $\mu=0$ but does not
exhibit a large-deviation Lifshitz mechanism.

\section{Third Law: Single Oscillator vs.\ Lattice}
\label{sec:third-law}
A common narrative for disordered oscillators contrasts a single soft-mode degree of 
freedom with an extended lattice. The folded-Gaussian model clarifies what is---and 
is not---a thermodynamic distinction.

\paragraph{Spectral origin of the power law.}
For $\mu=0$ the half-normal density is finite at $\omega=0$, $p_{FG}(0)>0$. In both 
the single oscillator and the lattice, the Bose kernel at $k_BT\ll\hbar\sigma$ reduces 
to its classical limit, so the low-temperature heat capacity is set entirely by the 
phase-space volume available to soft modes:
\begin{itemize}
\item \emph{Single oscillator:} phase space is the frequency interval $[0,k_BT/\hbar]$, 
giving $\langle C \rangle\propto T$ and $\langle S \rangle\propto T$.
\item \emph{Lattice:} the van Hove factor $\omega^{d-1}$ supplies an extensive 
shell volume $\propto (k_BT/\hbar)^d$, giving $\langle C\rangle\propto N T^d$ and 
$\langle S \rangle/N\propto T^d$ ($d\ge1$).
\end{itemize}
The lattice freezes \emph{faster} than the isolated oscillator for $d>1$; 
both freeze exponentially only when $\mu>0$, where $p_{\mathrm{FG}}(\omega)$ is gapped.

\paragraph{Third-law terminology.}
Strictly, Nernst's postulate requires only $\lim_{T\to0}
\langle S(T)\rangle =\text{const}$, and both single oscillator and lattice give 
$\langle S\rangle \to0$, so the third law is formally satisfied. 
The oft-quoted ``third-law violation'' for disordered phonons 
in the literature \cite{lifshitz1964,matsuda1970,nieuwenhuizen1987} is therefore not a 
residual entropy but a spectral property: the absence of an energy gap and a 
power-law freeze-out instead of Debye exponential decay.
In the present model this is driven by the \emph{folded-Gaussian kink tail}, 
$\bar\rho(\omega)\sim\omega^{d-1}$, rather than a Lifshitz tail in the strict sense.
The lattice freezes faster than the isolated oscillator for $d>1$, but both 
lack an exponential gap when $\mu=0$.

\paragraph{Summary table.}
Table~\ref{tab:scaling} therefore distinguishes regimes by spectral scaling 
rather than by entropy floors. The single oscillator and its lattice embedding 
differ in \emph{how fast} they freeze, not in whether they freeze.

\begin{table}[h]
\centering
\caption{Low-temperature scaling (folded-Gaussian oscillators). 
\textit{Note:} $p_{\mathrm{FG}}$ is defined in Eq.\eqref{eq:fg-pdf}. All cases 
satisfy $\langle S\rangle \to0$ strictly;  
$\mu>0$ gives exponential gap, $\mu=0$ removes the gap.}
\label{tab:scaling}
\begin{tabular}{l|l|c}
System & Low-$\omega$ weight & Low $T$ scaling of $\langle C\rangle $\\ \hline
Single osc., $\mu>0$ & $p_{\mathrm{FG}}(\omega;\mu,\sigma)$ 
as in Eq.\eqref{eq:fg-pdf}& $\propto \ee^{-\hbar\mu/k_BT}$ \\
Single osc., $\mu=0$ & $\sqrt{2/\pi}\sigma^{-1}\ee^{-\omega^2/(2\sigma^2)}$ & $\propto T$ \\
Lattice, $\mu>0$ & $\omega^{d-1}p_{\mathrm{FG}}(\omega;\mu,\sigma)$ & 
$\propto T^d \,\ee^{-\hbar\mu/k_BT}$ \\
Lattice, $\mu=0$ & $\omega^{d-1}\ee^{-\omega^2/(2\sigma^2)}$ 
& $\propto T^d$ (faster than single) \\
\end{tabular}
\end{table}

\section{Concluding Remarks and Discussion}
\label{sec:conclusions}
The statistical mechanics of a quantum harmonic oscillator with folded-Gaussian 
frequency disorder is studied in detail, extending from a single degree 
of freedom to an uncoupled lattice. The model is exactly solvable in integral form, 
admits controlled analytic limits, and cleanly isolates how soft-mode tails 
interact with thermodynamic phase space.

The core achievements are summarized in the following four points:
\begin{enumerate}
\item For a single oscillator with $\mu=0$, the half-normal tail produces a 
\emph{linear} low-temperature heat capacity $\langle C\rangle\propto T$ and 
entropy $\langle S\rangle\propto T$, 
not exponential suppression. The third law holds strictly; the soft mode merely slows 
freezing.
\item For $N$ independent oscillators the heat capacity is extensive, with relative 
sample-to-sample fluctuations $\propto N^{-1/2}$ (self-averaging). The disorder 
is thermodynamically benign in the thermodynamic limit.

\item Embedding in a $d$-dimensional lattice adds the van Hove factor 
$\omega^{d-1}$, converting the single-particle $T$ law into 
$\langle C\rangle\propto N T^d$. For $\mu=0$ the system lacks an energy gap 
and freezes as a power law. This is sometimes loosely called a ``Lifshitz-tail 
signature'' in the phonon literature, but strictly it is a \emph{soft-mode kink tail} 
from iid frequency folding, not a large-deviation Lifshitz mechanism.

\item The sharp-$\sigma$ cumulant expansion ($\mu\gg\sigma$) and the $\mu=0$ 
asymptotics exhaust the analytically tractable regimes; both limits are cross-checked 
against direct numerical quadrature of Eqs.\eqref{eq:avg-F}--\eqref{eq:avg-C}.
\end{enumerate}

The folded Gaussian is deliberately minimal: it guarantees positivity, keeps a 
Gaussian parent, and introduces a kink at $\omega=0$ that mimics the infrared weight 
of more complex disorder without Anderson localization. Its simplicity makes it a 
pedagogical null test for disorder--quantum-statistics effects, and a transparent 
contrast to truncated Gaussians or heavy-tailed frequency distributions. Future 
extensions could include spring-constant disorder (true Lifshitz tails), coupling 
to a bath, or entanglement entropy scaling across a folded-Gaussian interface.

%%%%%%%%%%%%%%%%%%%%%%%%%%%%%%%
\section*{Acknowledgement}
%%%%%%%%%%%%%%%%%%%%%%%%%%%%%%%
This work is supported by the National Natural Science Foundation 
of China under grant No. 12275138.

%%%%%%%%%%%%%%%%%%%%%%%%%%%%%%%
\section*{Data Availability Statement} 
%%%%%%%%%%%%%%%%%%%%%%%%%%%%%%%
This research makes no use of new data. 

%%%%%%%%%%%%%%%%%%%%%%%%%%%%%%%
\section*{Declaration of competing interest}
%%%%%%%%%%%%%%%%%%%%%%%%%%%%%%%
The author declares no competing interest.

%%%%%%%%%%%%%%%%%%%%%%%%%%%%%%%%%%%%%%%%%%

% ============================================================
% Bibliography
% ============================================================
%\bibliographystyle{unsrt}
%\bibliography{references}

\end{document}